\documentstyle[12pt]{article}

\begin{document}
\title{Exact Results of the Bit-String Model for Catastrophic Senescence.}
\author{T.J.P. Penna and S. Moss de Oliveira\\
Instituto de F\'{\i}sica, Universidade Federal Fluminense\\
Av. Litor\^anea s/n, Boa Viagem,\\
24210-340 Niter\'oi, RJ, Brazil\\
e-mail: tjpp@if.uff.br and suzana@if.uff.br}
\date{\today}
\maketitle
\begin{abstract}
We succeeded in obtaining exact results of the bit-string model of biological
aging for populations whose individuals breed only once. These results are in
excellent agreement with those obtained through computer simulations. In
addition, we obtain an expression for  the minimum birth needed to avoid
mutational meltdown.
\end{abstract}

\section{Introduction}

The aging is characterized by the loss of capabilities, in an irreversible way,
predominantly after the maturation, for the most of species. Many factors can
influence the aging process: biochemical processes, enviroment and so forth.
Recently, many efforts have been directed towards an evolutionary theory of
aging \cite{rose}-\cite{pb}. The fundamental question: Why people get old ? has
been recently replaced by:
Why is the survival probability of an individual reduced with advancing age ?
The reason is the latter is more suited to be answered with the
tools of dynamics of population and Mathematical Physics methods, as suggested
by Rose\cite{rose}. For a review of Monte Carlo
simulations in biological aging see, for instance, ref.\cite{bjp}. The theories of
evolutionary aspects of aging fall in two classes: optimality -- the optimal
life strategy is an emphasis on early
maturation that, due to physiological constraints, leads to a decrease in the life
span; and genetic explanations -- deleterious (bad) mutations for old age are subject to
a weaker selection that tends to accumulate them later in the life time.

Sometimes, aging effects are displayed in an amazing way, like in the so-called
catastrophic senescence.
It is a well known fact that semelparous individuals, i.e., those which breed
only once, usually present catastrophic senescence, that is, they die shortly
after reproduction\cite{pb}. Owing to its commercial value, the Pacific salmon
is one of the
most important examples. The catastrophic senescence has been studied in a
model of optimality  (Partridge-Barton model)\cite{pb,njan}. Nevertheless, if
we accept that natural selection prevents the manifestation of deleterious
mutations before procreation, it is easy to understand why this premature
senescence occurs: After reproduction, the individuum does not need to be
protected anymore, since it is not going to produce offspring any longer.

The bit-string model of life history, recently introduced\cite{bitstr,tjds},
specifically makes use of a balance between hereditary mutations and
evolutionary selection pressure. Computer simulations of the bit-string model
have shown that the catastrophic senescence can also be understood within the framework of mutation accumulation theory. Furthermore, it was possible to show that senescence is a direct result of the age of reproduction\cite{thoms}. Here, we
present analytical calculations which confirm
that the life strategy of {\bf reproducing only once} leads to the
catastrophic senescence. This paper is organized as follows: in section two, we briefly describe the bit-string model. In section three, we derive our analytical results and discuss them by comparison with computer simulations. Finally, in section four we present our conclusions.

\section{The Model}\par

Consider an initial population $N(t=0)$ of individuals each one
characterized by a word of $B$ bits. This word is related to the genetic code
of each individual, since it contains the information of when the effect of a
deleterious mutation will take place (bit set to one) - or not (bit set to
zero) - during the life of the individual. We can consider the time as a
discrete variable running from 1 to $B$ years. Hence, if at time $t=i$ the
$i^{th}$ bit in the word equals one, it is considered that the individual will
suffer the effects of a deleterious mutation in that year and all following years. The mutations are
hereditary in the sense that, at the birth of a new individual, they are transmitted to
this new individual except for $M$ bits ($M$ will be called mutation rate,
hereafter), {\em randomly chosen}. These mutations can correspond to either an
additional deleterious mutation or a cleaning of a deleterious mutation present at the parent's
genome.
We can interpret this new mutation as a point mutation (if we are considering
asexual reproduction) or a recombination of the parent's genome in sexual
reproduction.
An individual will die due to aging effects if the number of deleterious
mutations that were relevant up to its age is equal to a threshold $T$.

Competition between the individuals of a population - for food and space, for
example - are introduced taking into
account an age-independent Verhulst factor, which gives to each individual a
probability $x=(1-N_{tot}(t)/N_{max})$ of staying alive. $N_{max}$ means the maximum
size of the population.
Each surviving individual generates, at the reproduction age $R$, $b$
offsprings. As stressed in ref.\cite{salmao}, the fact of individuals breed
only once in their lifetimes is the fundamental ingredient to explain the
catastrophic senescence.

\section{Analytical Calculations and Results}

The relevant parameters in our calculations are:
\begin{itemize}
\item $B$ - genotype length, in bits;
\item $b$ - birth rate;
\item $R$ - the reproduction age;
\item $x$ - the Verhulst factor;
\item $M$ - number of deleterious mutations added at birth.
\item $T$ - threshold starting from which the mutations lead to death.
\end{itemize}
Let us assume that the system has reached an equilibrium state
(this hypothesis is plausible since computer simulations have
shown that it always occurs). Since the age distribution and the
population are stationary, we can consider the Verhulst factor as
constant. In this work, for the sake of simplicity, we assume only deleterious mutations
at birth with $T=1$, i.e., the
effects of just one deleterious mutation is enough to cause the death of an
individual. The extension for different values of $T$ is straightforward,
although we have learned from previous computer simulations results
that it is not important  for the catastrophic senescence explanation.

Let us consider one has
${\cal N}$
individuals with zero age, after the stationary state has been reached.
The probability of a deleterious mutation to appear at
a given age is $(M/B)$, {\it for ages below $R$}, since in this case we can be
sure that the parent has no mutations up to age $R$ ($T=1$) and the mutations
at birth are randomly chosen (we are considering the case $M << B$). For ages
beyond $R$, we must take into account
the whole parent's history, and this will be done later. First ignoring births, we can expect
that the number of individuals
with age 1 will be
\begin{equation}
 N_1 = {\cal N} x \left( 1 - {M \over B} \right) = {\cal N} x  \left( {B-M
\over B} \right).
\end{equation}
Analogously,
\begin{equation}
N_2 = N_1 x \left( 1 - {M \over  B - 1} \right) = {\cal N} x^2
{(B-M) \over B}{(B-M-1)\over (B-1)}
\end{equation}
is the number of individuals with age 2. For general age $k \le R$, one has
\begin{equation}
N_k = {\cal N} x^k  \left( \prod_{i=1}^k {B-i-M+1\over
B-i+1}\right).\label{nk0}
\end{equation}
Now, we take into account the reproduction of the individuals at
age $R$. Since each individuum produces $b$ offsprings, the number of
individuals with zero age will be $b$ times the number of age $R$ in the
previous step:
\begin{equation}
N_0 = b\; {\cal N}\;  x^R \;\left(\prod_{i=1}^R {B-i-M+1 \over B-i+1}\right)
\end{equation}

Using the above equation, with ${\cal N} = N_0$, one can obtain the stationary
state condition
\begin{equation}
b\, x^R   \prod_{i=1}^R {{B-i-M+1} \over {B-i+1}} = 1
\label{stationary}
\end{equation}
As we should expect, this condition does not depend on $k$ but only on the
global parameters $R,M,b$ and $B$. For $M=1$ it reduces to $bx^R(B-R)/B=1$.

For ages $k>R$, we must garantee that no mutation occurred between ages $R$
and $k$.
The probability that no mutation has occurred until age $k$, in the
$t^{\rm th} = 1, 2 \dots$ ancestor (parent, grand parent, etc.) is
(see eq.\ref{nk0}):
\begin{equation}
\left(\prod_{i=1}^k {{B-i-M+1} \over {B-i+1}} \right)^t,
\end{equation}
and this is a necessary condition for the individual to be still alive, to
be fulfilled for all ancestors.
However, except for $M=0$,
{\bf this term goes to zero as $t$ increases}. In this way,
in order to have the stationarity condition satisfied, we must have
\begin{equation}
N_{k>R} = 0.
\end{equation}
This feature of death after reproduction, as mentioned before,
is called catastrophic senescence. Now, it is evident that the only important
ingredient for this effect is to reproduce only once. Hereafter, any value of
$k$ is considered to be at most equal to $R$.

Using eq.(\ref{stationary}) to compute the value of $x$, we obtain the
following expression for the stationary age distribution:
\begin{equation}
N_{k} = {\cal N}
\left( \prod_{i=1}^k {{B-i-M+1} \over {B-i+1}}\right)
\left( {1\over b} \;\prod_{i=1}^R {{B-i+1} \over {B-i-M+1}}\right)^{k/R}
\label{Nk}
\end{equation}

Using the definition of $x$ we find from eqs.(3) and (5):
\begin{equation}
N_{tot} = \sum_{k=1}^R N_k = {N_{max}} \left\lbrack
1 - \left( {1\over b} \prod_{i=1}^R {{B-i+1} \over {B-i-M+1}}\right)^{1/R} \;
\right\rbrack
\label{ntot}
\end{equation}
The dependence of $N_{tot}$ with R is shown in Fig.1, and compared with
simulation results. The agreement is excellent.

>From eq.(\ref{ntot}) we can obtain a minimum birth rate, $b_{min}$, as a condition for avoiding
mutational meltdown\cite{Lynch} ($N_{tot}$=0):
\begin{equation}
b_{min} = \prod_{i=1}^{R} {{B-i+1} \over {B-i-M+1}},
\end{equation}
which equals $B/(B-R)$ for $M=1$.
Obviously, we obtain the condition $b_{min}=1$ for $M=0$.
This expression is important because it corroborates the results
of refs.\cite{americo} since it shows that the mutational meltdown {\em does
not} depend on the size of population - at least for asexual semelparous species -- in
contrast with earlier biological assumptions that this effect occurs only on
small populations.
We present in fig.2 some curves for $b_{min}$ as a function of $M$. For
comparison, we present some results from computer simulations. However, since
we have worked only with integer values of $b$, the theoretical results should
be considered as an upper limit. For large values of $b$ (typically $b>10$) and
high mutation rates ($M>3$), the number of time steps needed to reach the equilibrium is
very large. For a more precise determination of $b_{min}$, we tried an exponential
fit, for $N_{tot}$ vs. $t$ (for $10000<t<100000$), and we checked the sign of the
exponent to determine whether the population would eventually vanish. Since  the
study of dynamical aspects of this model are not into the scope of this work,
we present only the more reliable results.

The relevant quantities in aging studies are the survival rates $S_k$, i.e., the
probabilities of an individual to survive to the next age. They can be easily
obtained from eq.(\ref{Nk}):
\begin{equation}
S_k = {N_k \over N_{k-1}} =
{{B - k - M +1} \over {B - k +1}}
\left( {1\over b}  \prod_{i=1}^R {{B-i+1} \over {B-i-M+1}}\right)^{1/R}
\end{equation}
The survival rate for age $k$ gives the probability that
individuas alive at age $k-1$ survive and reach age $k$.
In fig.3, we compare this expression for different values of $b$ and $R$ with
computer simulation results. The abrupt jump to zero at $k = R$ is clearly seen.

\section{Conclusions}

Using simple statistical arguments, we have solved analytically the bit-string model, for
semelparous species as the Pacific salmon and mayflies.
Our results confirm those presented in ref.\cite{salmao}, where it is
claimed that the life strategy of just one breeding attempt leads to
catasthrophic senescence, according to the mutation accumulation theory. In
addition, we can obtain an analytical expression for the population size of
semelparous species as function of mutation rate and reproduction age (an
exponential decay for the same conditions of space and food), and the minimum
birth rate to avoid mutational meltdown. Our results also
agree with those found by Bernardes and Stauffer, suggesting that the
mutational meltdown is not an effect of small populations. Nevertheless, we
have found, by computer experiments, that the time required for a population
to vanish can be very large. In this sense, we suggest that it can be worth
studying the dynamical aspects of the bit-string model, for both semelparous
and iteroparous species.

As a final remark we  wish to point out that although dealing with
semelparous organisms, the extension for iteroparous organisms (which breed
repeatedly) is straightforward, but the corresponding expressions are too lengthy
to be useful.

This work was motivated by a question raised by Naeem Jan
about a previous paper.  We are grateful to him for his interest in our work.
One of us (TJPP) also is grateful to Silvio Salinas and Nestor Caticha for
encouragement, and the other (SMO) to P.M.C. de Oliveira for important suggestions.
This work is partially supported by Brazilian agencies CNPq,
CAPES and FINEP.

\section*{Figure Captions}

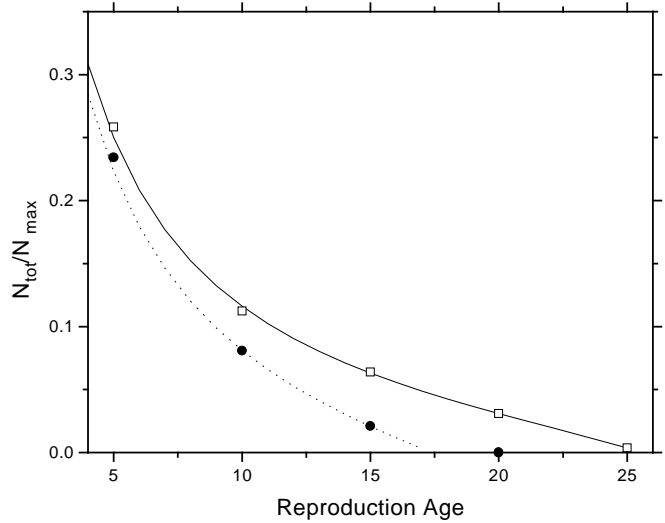
{\bf Fig.1} Number of individuals normalized by the factor $N_{max}$ after
15000 steps vs. reproduction age. We performed an average of the last 12000 steps
for ten different random numbers sequences. We consider $b=5$, $N_{max}=10^6$,
$M=1$ (solid line and open squares) and $M=2$ (dotted line and solid  circles).
The symbols correspond to the computer simulations and the lines correspond to
the exact results. Note the excellent agreement between theoretical predictions
and the computer simulations.

\bigskip
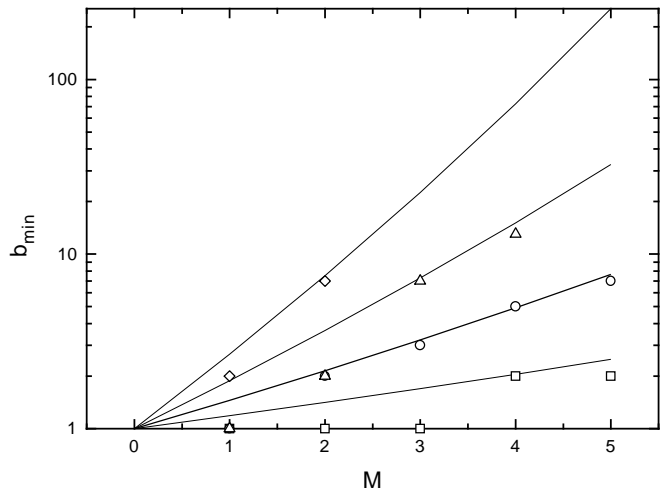
\noindent{\bf Fig.2} Minimum birth rate $b_{min}$ vs. mutation rate $M$. The
solid lines correspond to analytical results for $R=5,10,15,20$ from the bottom
to the top. The symbols correspond to computer simulations $R=5 (\Box), 10
(\bigcirc), 15 (\triangle)$ and $20 (\diamond)$. In the computer simulations,
we can only obtain the maximum integer value below which the population vanishes
(mutational meltdown). That is why these results are below the theoretical
results. We do not present results for larger values of $b$ because it would require a huge number
of time steps until the population vanishes.

\bigskip
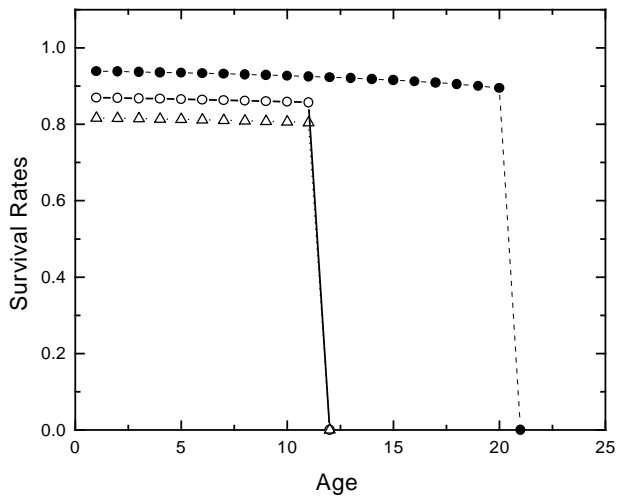
\noindent{\bf Fig.3} Survival rates vs. age. Here, we present the results for
$M=1$, $b=5$, $R=11$ ($\bigcirc$ and solid line), $R=20$ ($\bullet$ and dashed
line) and $b=10$, $R=11$ ($\triangle$ and dotted line). We followed the same
strategy of averaging as adopted in fig.1. Again, the
agreement between theory and computer experiment is remarkable.
\end{document}